\documentclass[conference]{IEEEtran}
\usepackage{cite}
\usepackage{graphicx}
\usepackage{amsmath}
\usepackage{amsfonts}
\usepackage{amssymb}
\usepackage[lined,boxed,ruled]{algorithm2e}
\begin{document}

\title{Social Decision Making with Multi-Relational Networks and Grammar-Based Particle Swarms}

\author{\authorblockN{Marko A. Rodriguez}
\authorblockA{CCS-3: Modeling, Algorithms, and Informatics\\
Los Alamos National Laboratory\\
Los Alamos, New Mexico 87545\\
Email: marko@lanl.gov}}

\maketitle

\begin{abstract}
Social decision support systems are able to aggregate the local perspectives of a diverse group of individuals into a global social decision. This paper presents a multi-relational network ontology and grammar-based particle swarm algorithm capable of aggregating the decisions of millions of individuals. This framework supports a diverse problem space and a broad range of vote aggregation algorithms. These algorithms account for individual expertise and representation across different domains of the group problem space. Individuals are able to pose and categorize problems, generate potential solutions, choose trusted representatives, and vote for particular solutions. Ultimately, via a social decision making algorithm, the system aggregates all the individual votes into a single collective decision.
\end{abstract}

\section{Introduction}

The social decision support system (SDSS) idea was first introduced in \cite{turoff:sdss2002} as a computer supported large group decision making system. An SDSS system allows a diverse set of individuals, in a computer mediated environment, to engage in a collaborative decision making process. The first SDSS publication formalized a social decision process model, an ontology for a collaborative discourse structure (i.e.~argumentation), and a means of determining the measure of confidence in a particular decision outcome (to accommodate fluctuating levels of participation). The SDSS was presented as a scaled version of the more common group decision support systems (GDSS). While most GDSSs are designed to support small groups of usually no more than 20 individuals \cite{gdss:hiltz1997}, the initial SDSS publication states that it can support on the order of thousands of participants. The SDSS extension presented in this article supports societal-scale decision making on the order of millions of participants via a partitioned problem space and trust-based social networks.\\

This article provides methods to support problem domains (i.e.~problem categories), dynamic power-structures via proxy-based representation (to accomadate fluctuating levels of participation), and a suite of vote aggregation algorithms (e.g.~direct democracy, representative democracy, dictatorship, etc.). This proposed SDSS extension provides a medium for individuals to express perceived problems, categorize problems, provide potential solutions, vote for particular solutions, choose representatives with respect to a problem domain, and ultimately yield a solution to a problem that is based on the aggregation of all local preferences--both individual preferences for solutions and voting representatives \cite{rodriguez:ci2004}. The single, unifying construct is a weighted multi-relational network that connects humans, problem space domains, proposed problems, and potential solutions in an ontologically constrained manner. This multi-relational network is the substrate which allows swarms of grammar-based particles to aggregate individual preferences into a final social decision. A grammar-based particle is a indivisible entity that propagates over a network according to some predefined finite state machine. These particles utilize the network topology to weight the influence of the various members of the group and in turn, rank the proposed solutions of a particular problem. Depending on the chosen grammar, different aggregation algorithms ranging from direct democracy to a dictatorship can be applied to the same underlying multi-relational network. Furthermore, as will be demonstrated, the proposed network ontology and particle swarm grammars can be extended to support collaborative discourse structures \cite{collab:turoff1999,turoff:sdss2002} without affecting the behavior of the vote aggregation algorithms presented. The major contribution of this research is the multi-functional framework along with the solutions it provides to previously articulated problems with the initial SDSS instantiation.\\

The next section will discuss the various stages of social decision making before formalizing the proposed multi-relational network ontology and grammar-based particle swarm algorithms.\\

\section{Social Decision Making}

\begin{figure}[h!]
  \begin{center}
    \includegraphics[width=0.45\textwidth]{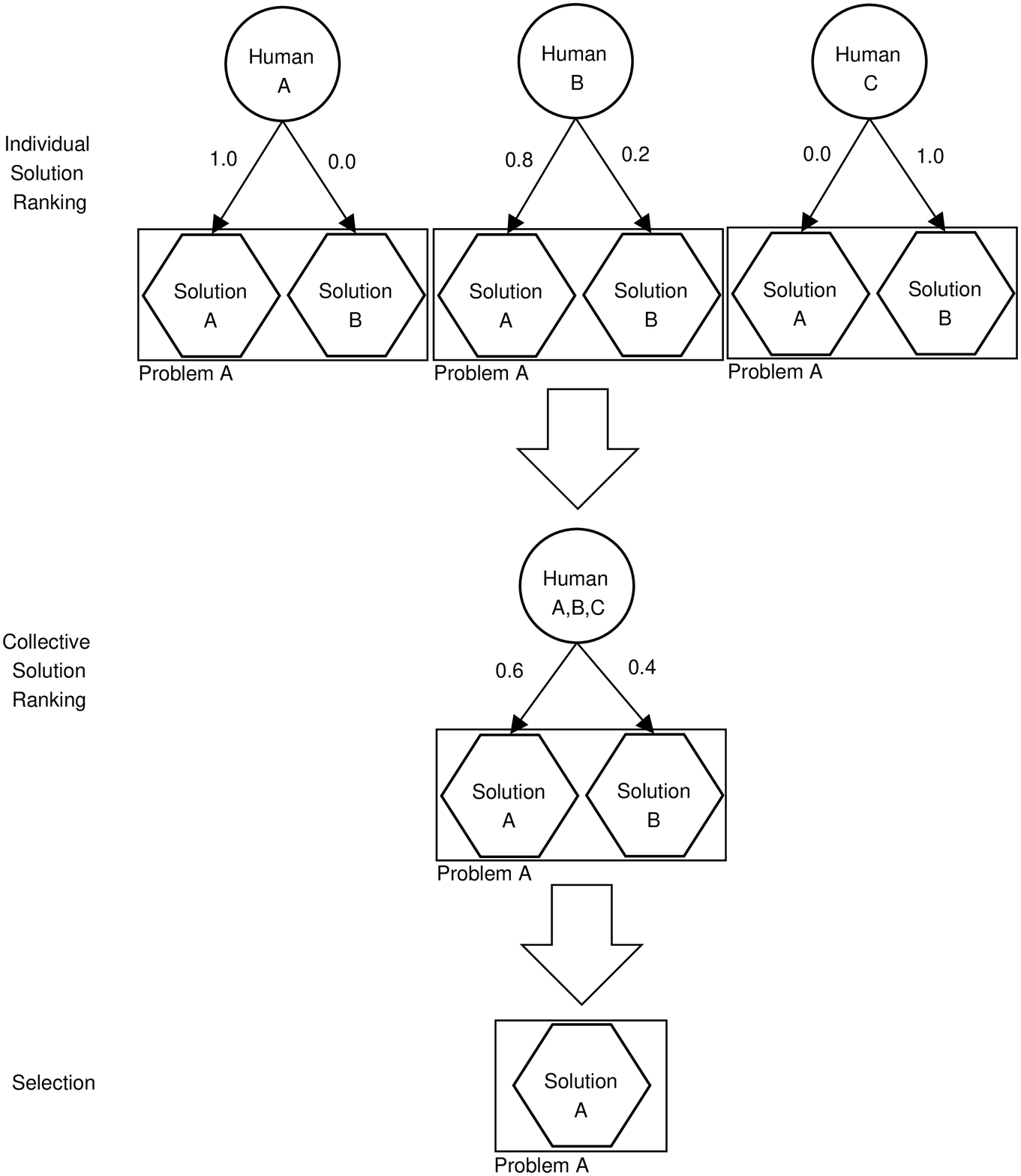}
    \caption{\label{fig:collective-decision} Three stage model of social decision making}
  \end{center}
\end{figure}

Social decision making is the process that takes every individual's local decisions and generates a single collective decision. This process can be deconstructed into three serial stages: \textit{individual solution ranking}, \textit{collective solution ranking}, and \textit{selection}. Figure \ref{fig:collective-decision} depicts the three stage social decision making process \cite{social:arrow1963}.\\

In stage one, individuals review solutions to a problem and rank them according to their subjective opinion. Therefore, the individual solution ranking algorithm exists internal to the individual. To go from individual rankings to a collective ranking, an algorithm is used to aggregate all the individual solution rankings into a single collective solution ranking. For example, in Figure \ref{fig:collective-decision}, the collective solution ranking algorithm is direct democracy (i.e.~one person/one vote) because each individual's solution ranking is weighted equally. Finally, given a collective solution ranking, a selection function is used to output a single final solution. For problems with numeric solutions (e.g.~a desired income tax), a final solution can be rendered via a weighted average of the ranked solutions. For problems with nominal, or categorical, solutions (e.g.~a potential candidate), the collective solution may be considered the one with the highest collective ranking. For example, in Figure \ref{fig:collective-decision}, the highest ranked solution, Solution $A$, is the collective solution.\\

The multi-relational network data structure formalized next makes explicit individual solution rankings. The particle swarm algorithms presented later use these individual solution rankings to derive a collective solution ranking. This collective solution ranking can then be used by a selection function to yield a single final solution to the originally posed problem.\\

\section{The Multi-Relational Network}

A multi-relational, or semantic, network is used to represent a heterogeneous set of entities (nodes) and a heterogeneous set of relationships (edges) \cite{sowa:semantic1991}. Therefore, there exist not only multiple entities of varying type, but also different ways, or semantics, by which these entities are connected. The \textit{ontology} for a particular multi-relational network defines the type of entities and relationships that can be instantiated in the network. This section will describe the social decision making ontology used throughout the remainder of this paper. Extensions can be made to the presented ontology without affecting the performance of particle swarm algorithms presented later. The proposed ontology contains four entities for implementing a social decision making system: \textit{humans}, \textit{domains}, \textit{problems}, \textit{solutions}, and their various relations.\\

The entities and their relations are represented by a directed, weighted, labeled network. Formally, this network is defined by the tuple $G = (N,W)$. For $G$, $N$ is the set of nodes in the network and $W$ is the set of weighted labeled edges such that $W \subseteq N \times N$, $w^{\lambda}_{n_i,n_j}$ connects node $n_i$ to $n_j$ according to the semantic label $\lambda$, and $w^{\lambda}_{n_i,n_j} \in \mathbb{R}$. Furthermore, $\lambda \in \Sigma^*$ where $\Sigma$ is some alphabet. The set $N$ is composed of four non-intersecting sets of heterogeneously typed nodes. The four sets are human nodes $H$, the domain nodes $D$, the problem nodes $P$, and the solution nodes $S$. Note that $H \cup D \cup P \cup S = N$ and $H \cap D \cap P \cap S = \varnothing$. These four sets and the description of their nodes are as follows:\\

\begin{itemize}
  \item $H \subseteq N$: the set of all humans participating
  \item $D \subseteq N$: the set of all domains used by humans to categorize problems and trust relationships
  \item $P \subseteq N$: the set of all problems
  \item $S \subseteq N$: the set of all solutions proposed for the problems in $P$\\
\end{itemize}

Each human node, $h_i \in H$, is associated with a personal collection of domain nodes used to categorize their social relationships and to categorize their problems. This idea is similar to `tagging' in collaborative classification systems \cite{tag:mathes2004,golder:collab2006}. The set of domain nodes for human $h_i$ are denoted $h_i(D)$ such that $D = \bigcup_{\forall i} h_i(D)$. Any particular domain $d_l$ of human $h_i$ is indexed as $h_i(d_l)$. Furthermore, $h_i(D) \cap h_j(D) = \varnothing \; : \; i \neq j$. Any problem $p_j$ has a collection of solution nodes associated with it, $p_j(S)$, such that $S = \bigcup_{\forall j} p_j(S)$. Furthermore, $p_i(S) \cap p_j(S) = \varnothing \; : \; i \neq j$. The remainder of this section will describe the graph-theoretic representations of humans and domains (the social space) and problems and solutions (the problem space).\\

\subsection{The Social Space}

The social space is the subset of $G$ that contains all humans, and the domains they use to classify problems and their various relationships, $H \cup D$. In the social space's most simplistic form, humans are connected to one another by generic trust edges, where trust is irrespective of the domain. With the inclusion of domains, it is possible to contextualize the type of trust one human has for another. This section will describe the various trust-based social network constructs.\\

\subsubsection{Social Trust Networks}

The relationships that people create between one another in the social space are trust relations for decision making. For example, human $h_0$ can make explicit his or her trust in human $h_1$. Trust relations form a social network that is used to weight the influence of various decision makers in the decision making process. The meaning of the trust relation will depend on the domain context and the individuals being connected. For example, trust could be a measure of similarity \cite{trust:ziegler2006}: human $h_0$ believes that human $h_1$ will make decisions that are in accord with their value system or $h_0$ perceives $h_1$ to possess comparable expertise. In other contexts, trust is based on differences in expertise. For example, if $h_0$ lacks expertise and believes that $h_1$ has expertise then $h_0$ believes $h_1$ will make a better decision. The degree of trust that $h_i$ has for $h_j$ can be represented by the conditional probability, $\mathsf{P}(A|B)$,

\begin{equation}
 w^{trusts}_{h_i,h_j} = \mathsf{P}(h_j \text{ is good} \; | \; h_i \text{'s knowledge of } h_j)
\end{equation}

Furthermore, given that humans are highly diverse (i.e.~multi-faceted, multi-skilled), trust relations tend to be domain specific, or context sensitive. This is the role of the set of domain nodes of an individual, $h_i(D)$. Human $h_0$ can trust that $h_1$ will make `good' decisions in the domain of $d_0$. Generally speaking,

\begin{eqnarray*}
 w^{trusts}_{h_i(d_l),h_j} = \\ &\mathsf{P}(h_j \text{ is good in } d_l \; \newline \\&| \; h_i \text{'s knowledge of } h_j \text{ in } d_l)
\end{eqnarray*}

As stated previously, the definition of `good' is left to the individual generating the trust-based relationship. There are two models of trust-based social networks to be described next.\\

\begin{itemize}
  \item \textit{single domain model}: trust has an undifferentiated context
  \item \textit{multiple domains model}: trust is context-sensitive\\
\end{itemize}

\subsubsection{\label{sec:sdm}Single Domain Model}

\begin{figure}[h!]
  \begin{center}
    \includegraphics[width=0.20\textwidth]{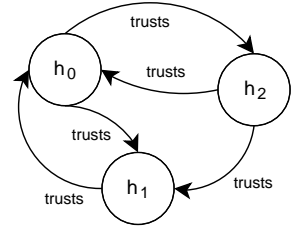}
    \caption{\label{fig:single-domain} A trust-based social network where trust is irrespective of context}
  \end{center}
\end{figure}

The simplest model, the single domain model, represents all trust relations and problem categories under a single domain context. Therefore, a network using this model does not actually require the domain construct. An example of a trust relation in the single domain model is $w^{\mathrm{trusts}}_{h_i,h_j}$. This weighted edge states that human $h_i$ trusts human $h_j$ to some degree, irrespective of the context. A context-insensitive social trust network is shown in Figure \ref{fig:single-domain}.\\

For social decision making systems with a focused problem space (i.e.~a single domain), the single domain model would be the most appropriate since the complexity of the implementation is minimal. However, for most social decision making systems, the problem space will tend to be composed of problems from various domains and therefore the trust that one delegates to another may need to be contextualized. For example, $h_0$ may trust human $h_1$ in domain $d_0$, but not in domain $d_1$. Therefore, in such situations, the single domain model is insufficient to represent the complexity of the social space.\\

\subsubsection{\label{sec:mrd}Multiple Domains Model}

The multiple domains model allows the individual to specify the context for which trust is given. Any human $h_i \in H$ is associated with a collection of unique domain nodes, $h_i(D)$. The purpose of domains is to allow individuals to categorize the type of trust relations they have with one another and to categorize the types of problems contained within the problem space (see section \ref{sec:ps}). Therefore, in this model, it is possible for $h_0$ to project a trust-based relation to $h_1$ in the domain $d_0$. Domain specific trust can be represented in many ways. The first representation is provided to help the reader to better understand this model as a transition from the single domain model presented previously.\\

\begin{figure}[h!]
  \begin{center}
    \includegraphics[width=0.40\textwidth]{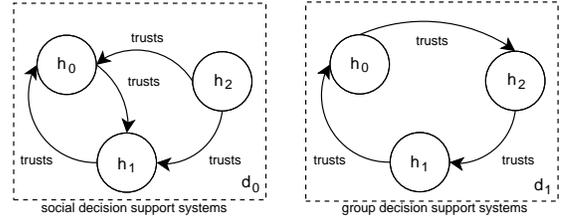}
    \caption{\label{fig:unrelated-domain} A representation of the multiple domains model}
  \end{center}
\end{figure}

For the first representation of the multiple domains model, it is possible to allow the group to create individual trust-based social networks for each domain of the problem space. This idea is depicted in Figure \ref{fig:unrelated-domain} where there exist two example domains, $d_0$ (social decision support systems) and $d_1$ (group decision support systems), and therefore, there exist two independent trust-based social networks.\\

It happens to be the case that, in Figure \ref{fig:unrelated-domain}, $d_0$ and $d_1$ are similar domains because much of the literature on social decision support systems refers to group decision support system research. Therefore it is likely that, because $h_0$ trusts $h_1$ in domain $d_0$, $h_0$ would also trust $h_1$ in domain $d_1$, to some degree. The idea of inferred trust through domain similarity cannot be made explicit in the above multiple domains model representation because domains are not related by any measure of similarity. To support domain similarity, the allowed relationships in the multiple domain model are as follows:\\

\begin{itemize}
  \item $w^{\mathrm{uses}}_{h_i,h_i(d_l)}$: human $h_i$ uses domain $h_i(d_l)$ to categorize entities
  \item $w^{\mathrm{trusts}}_{h_i(d_l),h_j}$: human $h_i$ trusts human $h_j$ in domain $d_l$
  \item $w^{\mathrm{similarTo}}_{h_i(d_l),h_i(d_m)}$: human $h_i$ believes that domain $d_l$ is similar to domain $d_m$\\
\end{itemize}

\begin{figure}[h!]
  \begin{center}
    \includegraphics[width=0.45\textwidth]{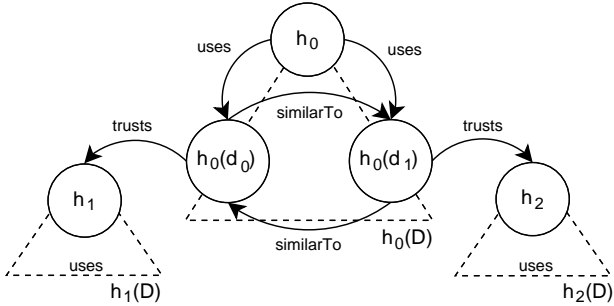}
    \caption{\label{fig:related-domain} A representation of the multiple related domains model}
  \end{center}
\end{figure}

The set of $h_i$'s personal domain nodes, $h_i(D)$, are related by the property of $\lambda = \mathrm{similarTo}$. For instance, $w^{\mathrm{similarTo}}_{h_0(d_0),h_0(d_1)}$ states that, human $h_0$ believes that domain $d_0$ is similar to $d_1$ by the amount dictated by the edge weight. By allowing the explicit representation of domain similarity, the number of trust-based edges between individuals can be reduced without losing necessary trust inference information. For instance, in Figure \ref{fig:related-domain}, if human $h_0$ believes that domain $d_1$ is similar to domain $d_0$, and if $h_0$ trusts $h_1$ within domain $d_0$, then it can be inferred that $h_0$ trusts $h_1$ within $d_1$ by an amount that is a function of $w^{\mathrm{similarTo}}_{h_0(d_1),h_0(d_0)}$.\\

The remainder of this paper will focus specifically on the more complicated multiple related domains model. Note that it is trivial to modify the forthcoming particle swarm algorithms to support the single domain model.\\

Finally, domain similarity and the creation of domains in the system can occur through text co-occurrence and the general collaborative tagging model \cite{tag:mathes2004}, respectively. The algorithms driving similarity generation and tagging are best defined by the specific system implementations and therefore, are outside the scope of this paper.\\

\subsection{\label{sec:ps}The Problem Space}

\begin{figure}[h!]
  \begin{center}
    \includegraphics[width=0.30\textwidth]{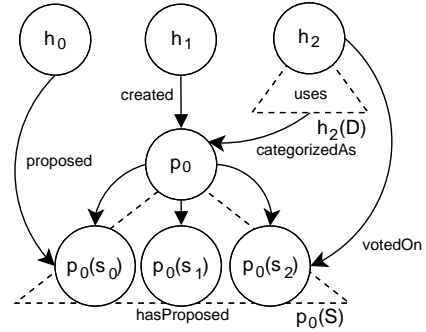}
    \caption{\label{fig:problem-space} The unification of the social and problem spaces}
  \end{center}
\end{figure}

The problem space is the set of all problems created and solutions proposed by the group, $P \cup S$. In the problem space, individuals can create and categorize problems and propose and vote on solutions. The solution nodes, $S$, represent the set of all proposed solutions to the problems facing the group. The set $p_j(S)$ is the set of all solutions proposed for problem $p_j$ such that $S = \bigcup_{\forall j} p_j(S)$. The unification of the problem space and the social space is depicted in Figure \ref{fig:problem-space}.\\

\begin{itemize}
  \item $w^{\mathrm{hasProposed}}_{p_j,p_j(s_m)}$: problem $p_j$ has proposed solution $s_m$
  \item $w^{\mathrm{created}}_{h_i,p_j}$: human $h_i$ created problem $p_j$ 
  \item $w^{\mathrm{categorizedAs}}_{h_i(d_l),p_j}$: human $h_i$ categorized problem $p_j$ as domain $d_l$ 
  \item $w^{\mathrm{proposed}}_{h_i,p_j(s_m)}$: human $h_i$ proposed solution $s_m$ for problem $p_j$ 
  \item $w^{\mathrm{votedOn}}_{h_i,p_j(s_m)}$: human $h_i$ voted on solution $s_m$ as the solution for $p_j$ \\
\end{itemize}

The problem space model presented thus far is only what is required to implement basic context specific decision making. Other problem space models do exist and have been reported in the literature. For instance, the argumentation problem space model of the SDSS (social decision support system) described in \cite{turoff:sdss2002} not only has problems (issues) and solutions (options), but also comments. Comment nodes allow individuals to remark on the proposed solutions to a problem. Comments have edges to solutions by way of the semantic projections $\lambda \in \{\mathrm{pro}, \mathrm{con}, \mathrm{neutral}, \mathrm{inquiry}\}$. Furthermore, comments can also have edges to other comments by means of the semantics of $\lambda \in \{\mathrm{opposing}, \mathrm{complementary}\}$. This argumentation problem space model allows users to communicate (argue) with one another for the purposes of creating consensus prior to collective decision making \cite{collab:turoff1999}.\\

The particle swarm algorithms to be described in Section \ref{sec:crd} are grammar-based algorithms that calculate functions only on certain subsets of the full network. This means that particle behavior is dictated by the $\lambda$ edges of the network. Therefore, the inclusion of more edge and node types will not affect these algorithms. For this reason, it is possible to extend the proposed ontology. Also, it is possible to include more particle swarm algorithms to take advantage of more node and edge types.\\

\section{\label{sec:crd}Collective Ranking with Particle Swarms}

Thus far, what has been presented, is a collective decision making ontology that determines the entities and relations of a particular multi-relational network data structure instantiation. Contained within any instantiation is a set of humans, their domains, their problems, and their solutions. It is understood, that via some interface, individuals are able to create problems, propose solutions, vote on solutions, create trust relations, instantiate domains, and modify domain similarities. That is, the individuals in the collective are able to directly affect the evolution of the multi-relational network data structure.\\

Of importance to collective decision making is determining the collective solution ranking for a particular problem (see Figure \ref{fig:collective-decision}). The results of the individual solution ranking algorithms internal to the individual are made explicit by means of the $\mathrm{votedOn}$ edges. In order to move from an individual preference to a collective preference, it is necessary to compute a collective solution ranking algorithm on the network. To do this, a general framework for aggregating perspectives is presented. A grammar-based particle swarm traverses the multi-relational network, identifying strong solutions as being those which have the most and shortest paths from the human nodes. By varying the parameters of the particle swarm, different collective solution ranking algorithms can be instantiated--from a dictator scenario to a pure direct democracy.\\

Section \ref{sec:gps} will focus on the the generic particle swarm shell which all particle swarm instantiations rely upon before discussing the particulars of the various collective solution ranking algorithms in Section \ref{sec:csra}. Finally, Section \ref{sec:cdra} will demonstrate how a particle swarm can be used to determine the domain of a problem.\\

\subsection{\label{sec:gps}The Generic Particle Swarm}

A particle swarm, as used in this context, is a collection of indivisible entities that propagate through a network in order to calculate a node rank distribution \cite{partpage:rodriguez2005}. In short, the more particles traverses a node, the more energy that node will receive. Furthermore, high energy nodes are considered highly ranked/valuable/important. Each particle swarm algorithm presented has a different method, as determined by their grammar, for biasing a particle's propagation and thus each calculate different nodal rankings.\\

Given a network $G$, a particle swarm can be used to determine the rank of a set of nodes relative to another set of nodes. These algorithms are sometimes called relative rank, or network influence, algorithms \cite{markov:white2003,partpage:rodriguez2005}. Within the context of collective decision making, for a particular problem $p_j$, it is desirable to rank each solution, $p_j(S)$, relative to a group of humans, $H$. In solution ranking, the humans compose the input node set, $I \subseteq H$, and the solutions to a problem $p_j$ form the output node set, $O = p_j(S)$. Furthermore, beyond solution ranking, other node rankings can be calculated.  For example, it may be desirable to rank each domain, $I = D$, relative to a particular problem, $O = p_j$. This is called domain ranking and is used to determine problem categorization (see section \ref{sec:cdra}).\\

The different particle swarm ranking algorithms presented in this paper all rely upon a few constructs. First, any particle $q_i \in Q$, is composed of two variables: a reference to its current node $c_i \in N$ and its current energy value $\epsilon_i \in \mathbb{R}$. These variables are itemized below for ease of reference:\\

\begin{itemize}
 \item $c_i \in N$: the current location of particle $q_i$
 \item $\epsilon_i \in \mathbb{R}$: the current energy content of particle $q_i$ \\
\end{itemize}

\begin{figure}[h!]
  \begin{center}
    \includegraphics[width=0.175\textwidth]{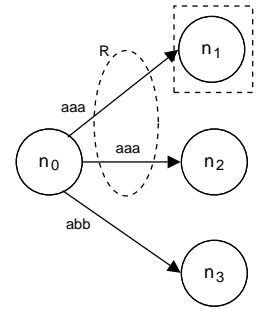}
    \caption{\label{fig:r-set} Edge traversal is dependent upon a subset of all outgoing edges of $c_i$}
  \end{center}
\end{figure}

A particle $q_i$ starts at its home node $c_i(t=0)$ and traverses an outgoing edge from its current node $c_i$ at each time step $t$. Note that  all particles initially start with the same energy value (i.e.~$\epsilon_i(t=0) = \epsilon_j(t=0) = 1.0 \; : \; i \neq j$). Graph traversal is a stochastic process that requires $c_i$'s outgoing edge weights to form a probability distribution. Therefore, the weights must be normalized to $1.0$. It is important to note that, depending on the particle swarm grammar, only certain edges can be traversed. For example, in Figure \ref{fig:r-set}, the set $R$ is the set of edges that the particle can traverse if the grammar states that only $\mathrm{aaa}$ edges are allowed. The function $\Theta(R) \rightarrow c_i(t+1)$ is a stochastic node selection function that returns the next node destination of particle $q_i$ given the probability distribution formed over the weight set $R$. In Figure \ref{fig:r-set}, if the particle is currently at node $n_0$, then node $n_1$ is the value returned by $\Theta(R)$ at $t+1$, $c_i(t)=n_0$ and $c_i(t+1)=n_1$.\\

Every node in the network has an associated energy, or activation, value that is initially equal to one another. This is denoted by the vector $\vec{e} \in \mathbb{R}^{|N|}$. This vector is indexed by the node name. Therefore, human $h_i$'s activation value is $\vec{e}_{h_i}$ and the current node of particle $q_i$'s energy value is $\vec{e}_{c_i}$. Each time a particle $q_i$ arrives at a node, it increments the node's activation value with its current energy value, $\epsilon_i$. This is represented in Eq. \ref{eq:addenergy}.

\begin{equation}
  \vec{e}_{c_i}(t+1) = \vec{e}_{c_i}(t) + \epsilon_i
  \label{eq:addenergy}
\end{equation}

Every time a particle leaves its current node, the particle decays its energy value by a global decay value $\delta \in \mathbb{R}$ as expressed in Eq. \ref{eq:decayenergy1} and \ref{eq:decayenergy2}.

\begin{equation}
 \epsilon_i(t+1) = (1-\delta) \epsilon_i(t)
  \label{eq:decayenergy1}
\end{equation}

such that,

\begin{equation}
 \epsilon_i(t) = (1-\delta)^{t} \epsilon_i(0)
  \label{eq:decayenergy2}
\end{equation}

Once a certain $k$ step has been reached or all nodal energy has decay to near zero, $\sum_{i=0}^{|Q|} \epsilon_i \approx 0.0$, the algorithm is complete and the desired output set's activation values can be normalized to $1.0$ to create a ranking over $O$ as demonstrated in Eq. \ref{eq:normalize}. The values in $\vec{e}(k+1)$ over the set $O$, denoted $\vec{e}_O$, is the output ranking.

\begin{equation}
 \vec{e}_{n_i}(k+1) = \frac{\vec{e}_{n_i}(k)}{\sum_{j=0}^{j < |O|} \vec{e}_{n_j}(k)} \; : \; n_i \in O, n_j \in O
  \label{eq:normalize}
\end{equation}

Algorithm \ref{alg:generic} provides the pseudo-code for ranking the entire network relative to itself ($I = O = N$). Algorithm \ref{alg:generic} is a discrete form of the iterative method for calculating the primary eigenvector of the full multi-relational network \cite{partpage:rodriguez2005}. This algorithm and its constructs form the foundation for the different collective ranking algorithms described next.\\

\incmargin{0.5cm}
\restylealgo{boxed}
\linesnumbered
\begin{algorithm}[h!]
\begin{footnotesize}
\Indp
  $\delta = 0.0; t=0$\;
  \CommentSty{\#iterate until some desired $k$-step}\;
  \While{$(t \leq k)$}{
    $t$++\;
   \CommentSty{\#distribute particles to the input set}\;
  \ForEach{$(n_i \in N)$}{
   $c_i = n_i; \epsilon_i = 1.0; q_i \neq \varnothing$\;
  }
  \While{$(\exists q_i \in Q \; : \; q_i \neq \varnothing)$}{
    \ForEach{$(q_i \in Q \; : \; q_i \neq \varnothing)$}{
      $\vec{e}_{c_i} = \vec{e}_{c_i} + \epsilon_i$\;
      $\epsilon_i = (1-\delta) \times \epsilon_i$\;
      $R = \bigcup_{\forall j,\lambda} w^{\lambda}_{c_i,n_j}$\;
      $c_i = \Theta(R)$\;
      \If{$(\epsilon_i \approx 0.0)$}{
       $q_i = \varnothing$\;
      }
    }
   }
  }
  \caption{\label{alg:generic}Generic particle swarm rank algorithm for all $\lambda$ edges}
\end{footnotesize}
\end{algorithm}
\decmargin{0.5cm}

\subsection{\label{sec:csra}Collective Solution Ranking Algorithms}

A collective solution ranking algorithm is the process that aggregates the different individual solution rankings into a single collective ranking. As will be demonstrated, there are many ways to aggregate individual solutions. The following sections will describe four collective solution ranking algorithms: direct democracy (DD), representative democracy (RD), dynamically distributed democracy (DDD), and dictatorship (D). Table \ref{tab:scenarios} outlines a few scenarios in which the various algorithms are most appropriate where `large ps' refers to a social decision support system with a large problem space (i.e.~many problems and problem domains). A double arrow is intended to signify an exaggeration of the appropriateness, e.g.~$\Uparrow$ signifies `most appropriate'. The reasons for the provided arrows are explained in the sections describing each algorithm.\\

\begin{table}[h!]
	\begin{footnotesize}
	\centering
		\begin{tabular}{|c|c|c|c|}
			\hline
					& \textbf{large ps}  & \textbf{expert domains} & \textbf{value domains} \\\hline\hline
			\textbf{DD}	& $\downarrow$	& $\Downarrow$	& $\uparrow$ \\\hline
			\textbf{RD}	& $\uparrow$	& $\uparrow$	& $\uparrow$ \\\hline
			\textbf{DDD}	& $\Uparrow$	& $\Uparrow$	& $\Uparrow$ \\\hline
			\textbf{D}	& $\Downarrow$	& $\uparrow$	& $\Downarrow$ \\\hline
		\end{tabular}
	\caption{\label{tab:scenarios}Scenarios for various collective solution ranking algorithms}
	\end{footnotesize}
\end{table}

\subsubsection{Direct Democracy Swarm}

Direct democracy embodies the idea of one person/one vote \cite{democracy:lijphart1984}. Every individual in the group is allowed to rank the solutions to a particular problem $p_j$ and the influence of all individual solution rankings is equal. If an individual $h_i$ does not vote, then $h_i$'s preference is left out of the collectively derived solution ranking of $p_j(S)$.\\

The direct democracy particle swarm algorithm works as follows. First, every individual in the group is provided a single particle, $I = H$. Second, every particle must take a $w^{\mathrm{votedOn}}_{h_i,p_j(s_m)}$ edge. If no such edge exists, then the particle destroys itself. Finally, if a particle is at a solution node, then, the particle destroys itself. This process is repeated until the solution ranking stabilizes. This is represented in Eq. \ref{eq:cosine} by the cosine similarity calculation of the solution set energy vector, $O = p_j(S)$, at different time steps.

\begin{equation}
 \frac{\vec{e}_{O}(t) \cdot \vec{e}_{O}(t+1)}{\left\|\vec{e}_{O}(t)\right\| \cdot \left\|\vec{e}_{O}(t+1)\right\|} \approx 1.0
 \label{eq:cosine}
\end{equation}

\begin{figure}[h!]
  \begin{center}
    \includegraphics[width=0.30\textwidth]{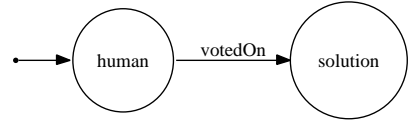}
    \caption{\label{fig:dd-fsm} Finite state machine describing the direct democracy solution ranking grammar}
  \end{center}
\end{figure}

A finite state machine representing the particle grammar is presented in Figure \ref{fig:dd-fsm}.\\

\begin{figure}[h!]
  \begin{center}
    \includegraphics[width=0.225\textwidth]{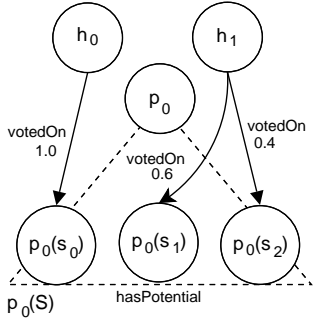}
    \caption{\label{fig:dd-viz} Repeated cycles ensures a proportionate energy distribution over an individual's solution rankings}
  \end{center}
\end{figure}

It is possible to allow an individual to vote on multiple solutions ($R \geq 1$). With multiple votes, the repeated cycling from human to solution as $t \rightarrow \infty$ will ensure that the particles traverse each human's voted on solutions the number of times proportionate to the weight value $w^{\mathrm{votedOn}}_{h_i,p_j(s_m)}$. For example, in Figure \ref{fig:dd-viz}, human $h_1$ has voted on two solutions with varying degrees of confidence. Each new time step $t$, it has a probability of taking one of the two $\mathrm{votedOn}$ edges as defined by the probability distribution over the edge weights. At $t=100$ the particles starting at $h_1$ will have arrived at solution $p_0(s_1)$ approximately 60 times and $p_0(s_2)$ approximately 40 times. Intuitively, if each individual is only allowed to choose one solution, then after $t=1$ the algorithm will have arrived at its stable solution set ranking.\\

For value based problems, the one person/one vote equality strategy of direct democracy may prove beneficial but, as the problem space grows in size, not all members of the collective will be able to participate in all decision making processes. Therefore, direct democracy is problematic in that those individuals who have not actively participated (i.e.~voted) have no influence in the collectively derived solution ranking. Moreover, the direct democracy strategy cannot identify expertise in a problem domain because without the trust-based social network, biasing the influence of different individuals is not possible.\\

\subsubsection{Representative Democracy Swarm}

The direct democracy strategy reduces the flexibility of the collective in that decision making processes cannot easily be parallelized. If each individual wishes to participate in the decision making of the group, then each individual must rank the solution set for each problem. Such serialization is not practical as the number of problems facing the group increases. With a large problem space, not all issues can be addressed by each and every individual. Therefore, in such situations, social decision making systems can rely on representation. Representation makes use of the domain-specific trust-based edges in the social space. Individuals that do not provide an individual solution ranking for a particular problem can delegate their decision making influence (i.e.~their particle) to a trusted representative.\\

Domain specific, or contextualized, representation refers to the notion that one can trust an individual in one domain, but not in another. For this reason, it is important to delegate an individual's decision making influence (i.e.~particle) to a representative with respect to the domain of the problem. Therefore, before delegation can occur, it is necessary to determine the problem's domain. Problem categorization is determined by way of the $\mathrm{categorizedAs}$ edge type that connects a domain to a problem. If human $h_0$ tagged problem $p_0$ with a $0.60$ $\mathrm{categorizedAs}$ edge from domain $h_0(d_1)$ and a $0.40$ edge from domain $h_0(d_2)$ then the particle has a 60\% chance of jumping to $h_0(d_1)$ and a 40\% chance of jumping to $h_0(d_2)$. Therefore, the weight $w^{\mathrm{uses}}_{h_0,h_0(d_2)}$ is $0.60$ and $w^{\mathrm{uses}}_{h_0,h_0(d_1)}$ is $0.40$. In general, for those who have provided a categorization of the problem, $\forall l \; w^{\mathrm{uses}}_{h_i,h_i(d_l)} = w^{\mathrm{categorizedAs}}_{h_i(d_l),p_j}$. It is important to realize that all $w^{\mathrm{uses}}_{h_i,h_i(d_l)}$ are determined at run-time, prior to particle diffusion. Furthermore, $w^{\mathrm{uses}}_{h_i,h_i(d_l)}$ weights are not directly manipulated by the individual, but instead indirectly through their $\mathrm{categorizedAs}$ edges. For those individuals that have not defined the domain of the problem (i.e.~have not provided a $\mathrm{categorizedAs}$ edge to the problem), a collectively generated domain is provided. The particle swarm method for determining the collectively derived domain of a problem is presented in Section \ref{sec:cdra}.\\

\begin{figure}[h!]
  \begin{center}
    \includegraphics[width=0.30\textwidth]{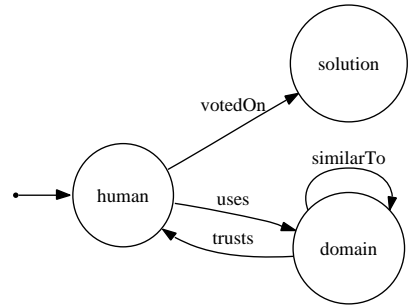}
    \caption{\label{fig:rd-fsm} Finite state machine describing the representative democracy solution ranking grammar}
  \end{center}
\end{figure}

A representative democracy particle swarm is an extension of the direct democracy swarm. First, every individual is supplied with a single particle. A particle, if it can, will take a $w^{\mathrm{votedOn}}_{h_i,p_j(s_m)}$ edge to an individual's chosen solution. If no such edge exists, then the particle will jump into the individual's domain network as specified by the probability distribution of the $w^{\mathrm{uses}}_{h_i,h_i(d_l)}$ edges. If there exists a $w^{\mathrm{trusts}}_{h_i(d_l),h_j}$ edge and the human $h_j$ has a $\mathrm{votedOn}$ edge to a particular solution to the problem (i.e.~$h_j$ is a voting representative), then the particle will take that edge. If not, then the particle will take a $w^{\mathrm{similarTo}}_{h_i(d_l),h_j(d_n)}$ edge in order to locate a voting representative in a related domain. If no such edges exist, then the particle destroys itself. Upon reaching a solution node, the particle will destroy itself. This process of distributing particles and propagating them from a human to a solution continues until the solution ranking has converged to a stable set of values (see Eq. \ref{eq:cosine}). The finite state machine of this grammar is presented in Figure \ref{fig:rd-fsm}.\\

\begin{figure}[h!]
  \begin{center}
    \includegraphics[width=0.225\textwidth]{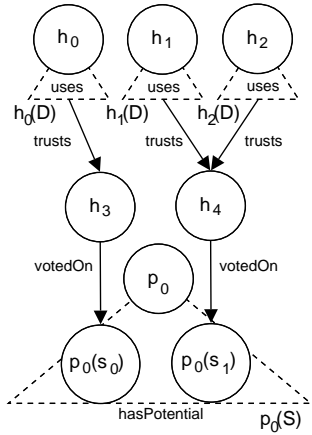}
    \caption{\label{fig:rd-viz} Tri-partite model of representative democracy}
  \end{center}
\end{figure}

Figure \ref{fig:rd-viz} demonstrates the tri-partite model of representative democracy.\\

It is important to note the energy decay parameter, $\delta$, in the representative democracy swarm. Intuitively, the further a particle must travel before it finds a solution node, the less that individual's vote is reflective of the individual's intentions. For example, if the particle goes from the human straight to a solution via a $\mathrm{votedOn}$ edge, then the intent of the individual is known and the effects of decay are less pronounced as only a single step is taken from source (human) to sink (solution). The human made explicit his or her desired solution to the problem. If the particle must move through the individual's network of domains to a representative, the more muddled the individual's perspective becomes. The further the particle travels from source to sink, the less explicit the perspective of the individual. Therefore, when the particle can find no solution to a problem, the individual has no model of that aspect of the problem domain (i.e.~it is not connected to the appropriate solutions or representatives) and therefore there is no way of knowing the individual's preference. To reflect this lack of information, the particle dies when too many steps have been taken, $\epsilon_i \approx 0.0$.\\

If trust is a measure of similarity of thought between two individuals within a particular domain, then one individual voting for two will reflect both individual perspectives--the participating individual's solution ranking is weighted twice. In such cases, representation is a form of social compression \cite{ddd:rodriguez2004} and it is this idea that allows for parallelization of the decision making efforts of the collective. If every individual in the group has provided a solution ranking for the problem, then representative democracy is identical to direct democracy as the particle is able to take a $\mathrm{votedOn}$ edge to a solution. On the other hand, if the domain of the problem is more expert based, individuals are weighted according to their perceived level of expertise. Individuals that abstain from participation are reducing the noise in the system (reducing poor individual solution rankings), and are, at the same time, ensuring that those individuals who are competent in the domain are weighted appropriately. Representative democracy is important in expert-based problem domains where non-expert participation decreases.\\

\subsubsection{Dynamically Distributed Democracy Swarm}

\begin{figure}[h!]
  \begin{center}
    \includegraphics[width=0.225\textwidth]{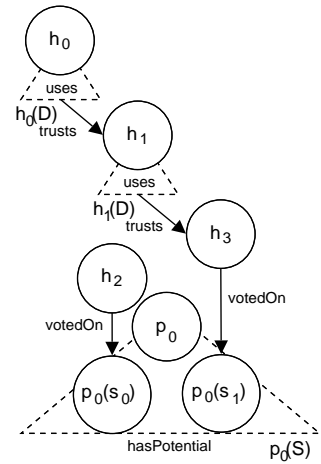}
    \caption{\label{fig:ddd-viz} Social network model of dynamically distributed democracy}
  \end{center}
\end{figure}

Dynamically distributed democracy (DDD) was first introduced in \cite{ddd:rodriguez2004} and is a slightly modified form of the representative democracy swarm. If every individual participates in the decision making process by providing a solution ranking, then DDD is identical to direct democracy. However, as individuals abstain from voting, DDD models the representative form of democracy in which an individual can delegate their decision making influence to a representative. Unlike representative democracy, that representative is not required to be a participating, or voting, individual. Therefore, representation is recursive as demonstrated in Figure \ref{fig:ddd-viz}.\\

If individual $h_i$ has not voted on a particular solution to a problem, then $h_i$'s particle is distributed to some trusted neighbor, $h_j$, within the domain of the problem. If that individual has not voted, then the particle is distributed to $h_j$'s trusted neighbor. So on and so forth until the particle can find an individual that has voted on a solution to the problem. DDD is a recursive version of the representative democracy swarm where one's representative may be steps away in the trust-based social network. Therefore, DDD is well suited for trust-based networks that are not bi-partite and are recurrent.\\

In expert based domains, $h_i$'s expertise is identified not solely by $h_i$'s in-degree, but also by the in-degree of the individuals that trust $h_i$. This recursive definition of expertise is analogous to the PageRank algorithm's calculation of web-page prestige \cite{google:brin1998} and may be a better measure of individual expertise \cite{social:wasserman1994}. DDD has been shown, in simulation, for value-based problems, to be able to weight the active voters appropriately such that any set of voting individuals rank the solution set $p_j(S)$ as if every member of the collective had participated. The weighted voters are said to form a holographic model of the whole population \cite{ddd:rodriguez2004}.\\

\subsubsection{Dictator Swarm}

The dictator swarm is the final collective solution ranking algorithm discussed by this paper and is at the extreme end of the representative forms of democracy. In direct democracy, every individual is only a representative of themselves. In a dictatorship, one individual is a representative of every other individual. Note that DD, RD, and DDD are dictatorships when only one individual of the group provides an individual solution ranking.\\

The dictator of a social group can be derived using various social network metrics such as eigenvector centrality, betweenness centrality, in-degree, etc. \cite{social:wasserman1994}. The dictator swarm rule is different than the direct democracy swarm in that if a particle ever finds itself at an individual that is not the dictator, it destroys itself. This simple method ensures that only the dictator is ranking the problem's solution set, and therefore, the collective solution ranking is identical to the dictator's individual solution ranking as demonstrated in Eq. \ref{eq:dictator}.

\begin{equation}
 \forall_m \; \vec{e}_{p_j(s_m)} = w^{\mathrm{votedOn}}_{h_i,p_j(s_m)} : h_i \; \text{is the dictator}
 \label{eq:dictator}
\end{equation}

\subsection{\label{sec:cdra}Categorizing a Problem}

\begin{figure}[h!]
  \begin{center}
    \includegraphics[width=0.4\textwidth]{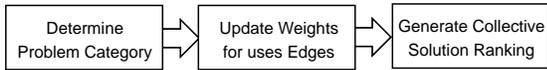}
    \caption{\label{fig:domain-ranking-stages} Problem category determined prior to collective solution ranking}
  \end{center}
\end{figure}

What has been presented thus far is a model of ranking solutions to a problem relative to a collection of humans. It is also possible for particle swarms to generate collective rankings of other entities in the network. In this section, a swarm grammar is introduced that is able to rank the collection of all domains, $D$, relative to a particular problem $p_j$. The purpose of a collective domain ranking algorithm is to determine, at run-time, the weights for the $\mathrm{uses}$ edges that connect a human node to a particular domain, $w^{\mathrm{uses}}_{h_i,h_j(d_l)}$. The importance of this is to ensure that particles propagate to representatives with respect to the problem domain. In order to do this, it is necessary to calculate the domain of the problem and therefore, determine what weights to give the $\mathrm{uses}$ edges as demonstrated in Figure \ref{fig:domain-ranking-stages}.\\

\begin{figure}[h!]
  \begin{center}
    \includegraphics[width=0.30\textwidth]{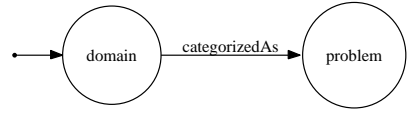}
    \caption{\label{fig:dd-domain-fsm} Finite state machine describing the direct domain ranking grammar}
  \end{center}
\end{figure}

\begin{figure}[h!]
  \begin{center}
    \includegraphics[width=0.30\textwidth]{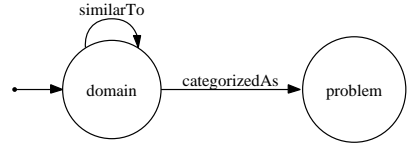}
    \caption{\label{fig:rd-domain-fsm} Finite state machine describing a recursive domain ranking grammar}
  \end{center}
\end{figure}

\begin{figure}[h!]
  \begin{center}
    \includegraphics[width=0.35\textwidth]{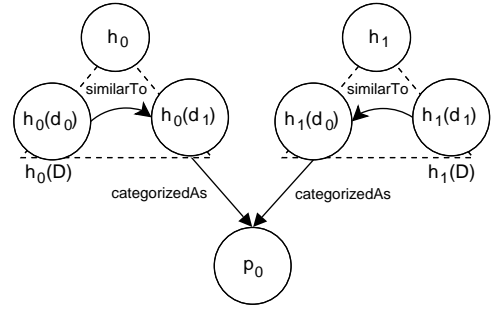}
    \caption{\label{fig:domain-ranking} Recursive domain ranking model}
  \end{center}
\end{figure}

In collective domain ranking, the $\mathrm{categorizedAs}$ edge is analogous to the $\mathrm{votedOn}$ edge of the solution ranking algorithms. Also, the $\mathrm{similarTo}$ edges are synonymous with the $\mathrm{trusts}$ edges of the representative forms of the solution ranking swarms. The input set for collective domain ranking is the set of all domains, $I = D$. The output set is the set of all domains that have a $\mathrm{categorizedAs}$ projection to problem $p_j$, $O = \bigcup_{\forall i,l} h_i(d_l) \; : \; \exists w^{\mathrm{categorizedAs}}_{h_i(d_l),p_j}$. The general rule is that a particle can only take a $\mathrm{categorizedAs}$ to problem $p_j$. In the recursive rank form, if no such edge exists then the particle must take a $\mathrm{similarTo}$ edge. Anytime a particle becomes stuck, it destroys itself, $q_i = \varnothing$. The direct domain ranking grammar is presented in Figure \ref{fig:dd-domain-fsm} and the recursive domain ranking grammar is presented in Figure \ref{fig:rd-domain-fsm}. A visualization of a recursive ranking of problem $p_0$ is presented in Figure \ref{fig:domain-ranking}. The direct form would not allow the the particle to take $\mathrm{similarTo}$ edges.\\

The normalization function to generate the $\mathrm{uses}$ weights set for particle entry into human $h_i$'s domain network is defined in Eq. \ref{eq:domainuses}.

\begin{equation}
	w^{\mathrm{uses}}_{h_i,h_i(d_l)} = 
	\begin{cases}
	  \frac{\sum_{x=0}^{x < |H|} \vec{e}_{h_x(d_l)}(k)}{\sum_{x=0}^{x < |H|} \sum_{y=0}^{y < |h_x(D)|} \vec{e}_{h_x(d_y)}(k)} & \text{if} \; \exists w^{\mathrm{categorizedAs}}_{h_x(d_l),p_j}\\
	  0 & \text{otherwise}
	\end{cases}
	\label{eq:domainuses}
\end{equation}

If the individual has provided their own categorization of the problem then, as stated previously, Eq. \ref{eq:domainuses2} provides the $\mathrm{uses}$ weight.

\begin{equation}
	w^{\mathrm{uses}}_{h_i,h_i(d_l)} = 
	\begin{cases}
	  w^{\mathrm{categorizedAs}}_{h_i(d_l),p_j} & \text{if} \; \exists w^{\mathrm{categorizedAs}}_{h_i(d_l),p_j}\\
	  0 & \text{otherwise}
	\end{cases}
	\label{eq:domainuses2}
\end{equation}

Note that direct democracy and dictatorships don't require problem categorization since proxy representation is never used (i.e.~the $\mathrm{uses}$ edges are not included in their grammars).\\

\section{Selecting a Final Outcome}

As originally presented in Figure \ref{fig:collective-decision}, the final stage of collective decision making is the selection function. The selection function takes the collective solution ranking, $\vec{e}_{p_j(S)}$, and generates a single collective decision outcome, $s_m \in p_j(S)$. Depending on the nature of the problem, different selection functions are more appropriate than others. For continuous numeric solution sets, averaging can be used. In experiments where participants guess how many beans are in a large jar, averaging individual guesses turns out to be a near-optimal strategy \cite{bean:treynor1987,ci:watkins2005}. For a nominal, or categorical, solution set, the `highest ranked wins' rule can be used. Various selection algorithms can be created to generate a single solution from a collective solution ranking. The breadth of potential selection functions is out of the scope of this paper.\\

\section{Conclusion}

\begin{figure}[h!]
  \begin{center}
    \includegraphics[width=0.49\textwidth]{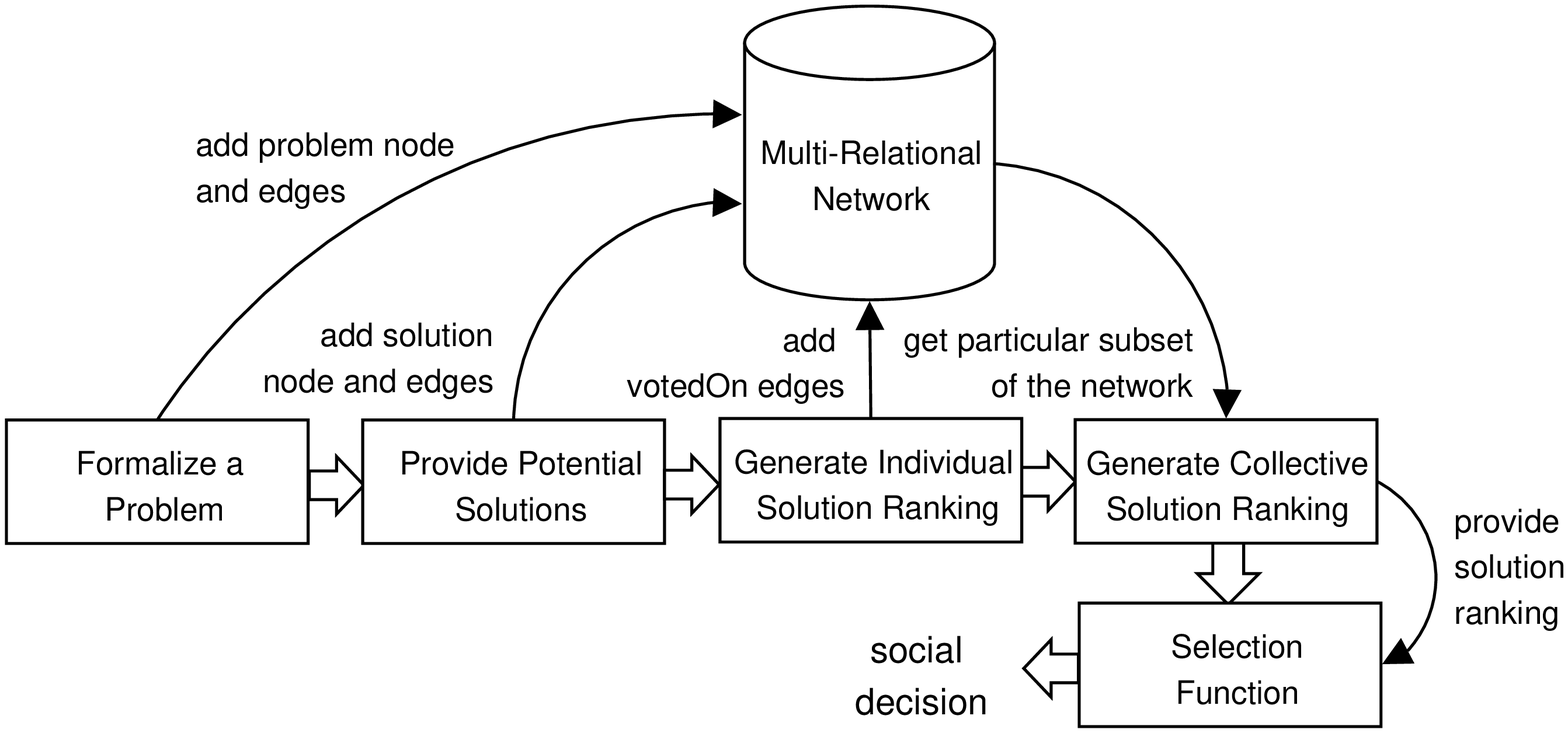}
    \caption{\label{fig:system-outline} Social decision making with the multi-relational network}
  \end{center}
\end{figure}

Figure \ref{fig:system-outline} represents the social decision making stages and their relationship to each other and the multi-relational network. Via a user interface, individuals are able to formalize problems and potential solutions for the various posed problems. Problem and solution formalization adds the appropriate nodes and edges to the multi-relational network. Next, individuals make explicit the results of their internal individual solution ranking algorithm by adding $\mathrm{votedOn}$ edges to the multi-relational network. A collective solution ranking algorithm is calculated with respects to a particular problem. The result of the collective solution ranking algorithms is inputed into a selection function which returns a single final solution--the social decision. Note that social network trust creation and problem categorization are not represented in Figure \ref{fig:system-outline}.\\

The presented social decision making system framework is a scalable solution for societal-scale decision support systems where ad hoc representative power structures are required to handle fluctuating levels of participation across the various problem domains of the problem space. The inclusion of discourse relational structures and supporting particle swarm algorithms provides the complete societal-scale decision support system as articulated by \cite{turoff:sdss2002,rodriguez:ci2004}.\\

\section*{Acknowledgments}

Carlos Gershenson, Daniel J. Steinbock, Francis Heylighen, Jennifer H. Watkins, and Johan Bollen have all contributed via extended discussions on the issues concerning this topic. Funding was made possible thanks to a GAANN Fellowship from the U.S. Department of Education, the Fonds voor Wetenschappelijk Onderzoek - Vlaanderen, and Los Alamos National Laboratory.\\.

        \bibliographystyle{IEEEtran}
        \bibliography{grammardecision}

\end{document}